\documentstyle[aps,pra,epsfig,twocolumn]{revtex} 

\def\be{\begin{equation}}
\def\ee{\end{equation}}
\def\bea{\begin{eqnarray}}
\def\eea{\end{eqnarray}}
\def\bma{\begin{mathletters}}
\def\ema{\end{mathletters}}
\def\C{\hbox{$\mit I$\kern-.6em$\mit C$}}

\begin{document}
\draft

\title{Optimal Conclusive Discrimination of Two Non-orthogonal Pure Product Multipartite States Locally}

\author{Yi-Xin Chen and Dong Yang}

\address{Zhejiang Institute of Modern Physics and
Department of Physics, Zhejiang University, Hangzhou 310027, P.R. China} 

\date{\today}

\maketitle

\begin{abstract}
We consider one copy of a quantum system prepared in one of two non-orthogonal pure product states of multipartite distributed among separated parties. We show that there exist protocols which achieve optimal conclusive discrimination by means of local operations and classical communications(LOCC) as good as by global measurements. Also, we show a protocol which minimizes the average number of local operations. Our result implies that two pure product multipartite states might not have the non-local property although more than two can have.  
\end{abstract}


\narrowtext

It is a key point that it is impossible to discriminate perfectly between non-orthogonal quantum states if only one copy is provided. Generally, orthogonal states may be distinguished with certainty by means of global operations since quantum information of orthogonality may be encoded in entanglement which may not be extracted by LOCC operations. Recently, Walgate et.al. \cite{Walgate} demonstrated that any two orthogonal multipartite pure states can be distinguished perfectly by only LOCC operations. Virmani et.al. \cite{Virmani} utilized their result to show that optimal discrimination of two non-orthogonal pure states can also be obtained by LOCC in the sense of inclusive discrimination. They also mentioned conclusive discrimination of certain regimes. Bennett et.al. \cite{Bennett} demonstrated that there exist bases of orthogonal product pure states which can not be locally reliably distinguished despite the fact that each state in the basis contains no entanglement. So product pure states may also have non-local property. In this note, we consider the problem of conclusive discrimination of two non-orthogonal product pure states by LOCC operations and show that the optimal disrimination achieved by global measurements can also be obtained by LOCC. This may implies that two product pure multipartite states might not have non-local property.

The problem of identifying two non-orthogonal states has been considered by Ivanovic \cite{Ivanovic}, Dieks \cite{Dieks}, Peres \cite{Peres}, Jaeger and Shimony \cite{Jaeger} by global measurements. Since we utilize their results especially Jaeger and Shimony's, we recall their results in brief and point out some properties which is implicit in their results. Suppose one copy of quantum state prepared as $|p>$ or $|q>$, generally non-orthogonal, occurring with probabilities $r$ and $s$ respectively, where $r+s=1$. The conclusive discrimination means that our measurement on the copy gives three outcomes which allow us to determine the prior state is $|p>$ or $|q>$ with certainty or "don't know". The optimization of conclusive discrimination is to obtain the maximal probability of decisive outcomes.

The procedures suggested by Dieks \cite{Dieks}, Peres \cite{Peres}, Jaeger and Shimony \cite{Jaeger} are the following: prepare an auxiliary system in a state $|s_{0}>$ and choose a unitary evolution which yields 
\bea
|s_{0}>|p> \longrightarrow \alpha |s_{1}>|p_{1}>+\beta |s_{2}>|p_{2}>, \nonumber\\
|s_{0}>|q> \longrightarrow \gamma |s_{1}>|q_{1}>+\delta |s_{2}>|q_{2}>, 
\eea
where $<p_{1}|q_{1}>=0, <s_{1}|s_{2}>=0$ and $|q_{2}>$ is identical with $|p_{2}>$ except for a phase factor. The measurement on the auxiliary system gives two outcomes: if $|s_{1}>$ occurs, a measurement between $|p_{1}>$ and $|q_{1}>$ gives the prior state $|p>$ or $|q>$ with certainty; if $|s_{2}>$ occurs, we get the outcome "don't know". The aim is to maximize the probability $P=r|\alpha|^{2}+s|\gamma|^{2}$ with the constraint of the procedures.

Without loss of any generality, $r \ge s$, Jaeger and Shimony's optimal result is:
If $\sqrt{\frac{s}{r}} \ge |<p|q>|$,
then
\bea
|\beta|^{2} &=& |<p|q>|\sqrt{\frac{s}{r}},\nonumber\\
|\delta|^{2} &=& |<p|q>|\sqrt{\frac{r}{s}},\nonumber\\
P_{max} &=& 1-2\sqrt{rs}|<p|q>|.
\eea  
If $\sqrt{\frac{s}{r}} \le |<p|q>|$, 
then
\bea
|\beta|^{2} &=& |<p|q>|^{2},\nonumber\\
|\delta|^{2} &=& 1,\nonumber\\
P_{max} &=& r(1-|<p|q>|^{2}).
\eea

In the above procedures, we notice that the outcome of "don't know" occurs with probability $P_{E}=r|\beta|^{2}+s|\delta|^{2}$, resulting from two states with equal probability in the former case while with unequal probability in the latter case. At first glance, they may be incompatible, but we will show that indeed they are compatible. Let's firstly express $P_{E}$ explicitly in respective case. Once the error occurs, in the former case, the error results from $|p>$ with probability 
\be
P_{|p>|E}= \frac{r|\beta|^2}{P_{E}} = \frac{1}{2},
\ee
from $|q>$ with 
\be
P_{|q>|E}= \frac{s|\delta|^2}{P_{E}}= \frac{1}{2};
\ee 
in the latter case, the error from $|p>$ with
\be
P_{|p>|E}= \frac{r|\beta |^2}{P_{E}}=\frac{r|<p|q>|^{2}}{P_{E}},
\ee
from $|q>$ with 
\be
P_{|q>|E}= \frac{s|\delta|^2}{P_{E}}=\frac{s}{P_{E}}.
\ee

Now we apply the result to two non-orthogonal pure product bipartite states. Suppose Alice and Bob are provided with one pure product state, $|u>=|u_{1}>|u_{2}>$ or $|v>=|v_{1}>|v_{2}>$ with probabilities $r$ and $s$ respectively, where $r+s=1$. Each half is available to one of them who are separated from each other and can communicate classical information only. Their aim is to identify the shared product state optimally in the sense of conclusive discrimination by means of local operations and classical communications. Our protocol is that Bob carries out the procedures which distinguish $|u_{2}>$ from $|v_{2}>$ with corresponding probabilities $r$ and $s$ optimally, while Alice performs those which distinguish $|u_{1}>$ from $|v_{1}>$ with conditional probabilities in which case Bob gets the error outcome even without obtaining information about Bob's outcome, or vice versa. The two protocols are equivalent. We set $|<u_{1}|v_{1}>|\ge |<u_{2}|v_{2}>|$ and $r\ge s$ with no loss of generality.

If 
\be
\sqrt{\frac{s}{r}} \ge |<u_{2}|v_{2}>|,
\ee
then
\be
\sqrt{\frac{s}{r}} \ge |<u|v>|,
\ee
and the optimal discrimination by global operation is 
\be
P^{G}=1-2\sqrt{rs}|<u|v>|.
\ee
Now by local operations, Bob first discriminates between $|u_{2}>$ and $|v_{2}>$ with $r$ and $s$ respectively. His error probability is 
\be
P_{E2}=2\sqrt{rs}|<u_{2}|v_{2}>|.
\ee
and when the error outcome occurs, the conditional probabilities from two prior states are equal. That is
\be
P_{|u_{2}>|E2}=P_{|v_{2}>|E2}=\frac{1}{2}.
\ee
Alice then discriminates between $|u_{1}>$ and $|v_{1}>$ with equal probabilities. Her error probability is 
\be
P_{E1|E2}=|<u_{1}|v_{1}>|.
\ee
So the total error is 
\be
P_{E}=P_{E2}P_{E1|E2}=2\sqrt{rs}|<u|v>|.
\ee
and the probability of successful discrimination is 
\be
P^{L}=1-P_{E}=1-2\sqrt{rs}|<u|v>|,
\ee
the same as that achieved by the global scheme. That means it is optimal.

If  
\be
\sqrt{\frac{s}{r}} \le |<u_{2}|v_{2}>|,
\ee 
then 
\be
P_{E2}=r|<u_{2}|v_{2}>|^2+s.
\ee 
and 
\bea
P_{|u_{2}>|E2} &=& \frac{r|<u_{2}|v_{2}>|^{2}}{P_{E2}}, \nonumber\\
P_{|v_{2}>|E2} &=& \frac{s}{P_{E2}}, \nonumber\\
P_{|u_{2}>|E2} & \ge & P_{|v_{2}>|E2}.
\eea
Now it's Alice's turn to discriminate between $|u_{1}>$ and $|v_{1}>$ with probabilities $r'=\frac{r|<u_{2}|v_{2}>|^{2}}{P_{E2}}$ and $s'=\frac{s}{P_{E2}}$.

If
\be
\sqrt{\frac{s'}{r'}}\ge |<u_{1}|v_{1}>|,
\ee
then
\be
\sqrt{\frac{s}{r}}\ge |<u|v>|,
\ee
and the global scheme gives the optimal discrimination  
\be
P^{G}=1-2\sqrt{rs}|<u|v>|.
\ee
In LOCC protocol, Alice error probability is 
\be
P_{E1|E2}=2\sqrt{r's'}|<u_{1}|v_{1}>|.
\ee
So the total error probability is 
\be
P_{E}=P_{E2}P_{E1|E2}=2\sqrt{rs}|<u|v>|,
\ee 
so the probability of decisive outcomes is also equal to $P^{G}$.

If 
\be
\sqrt{\frac{s'}{r'}}\le |<u_{1}|v_{1}>|,
\ee
then 
\be
\sqrt{\frac{s}{r}}\le |<u|v>|,
\ee
and the optimal discrimination by global scheme is 
\be
P^{G}=r(1-|<u|v>|^{2}).
\ee
By means of LOCC, Alice error probability is 
\be
P_{E1|E2}=s'+r'|<u_{1}|v_{1}>|^{2}.
\ee
So the total error probability is 
\be
P_{E}=P_{E2}P_{E1|E2}=s+r|<u|v>|^{2},
\ee
and $P^{G}$ is also achieved by LOCC.

It is explicit that we can achieve the same optimal discrimination by varying the order of operations performed by Alice and Bob. Further, it is easy to generalize to two non-orthogonal pure product multipartite states and also to achieve the optimal discrimination same as the global scheme. As to two pure product tripartite states, we can group bipartite as one and the third as another and perform the optimal discrimination between two parties. This can be realized by local discrimination, further the discrimination between the first two can also be realized locally. That completes our proof.

Our protocol implies that there exist many different local procedures that can achieve the same optimal discrimination. Among them, it is straightforward that we can select the protocol which minimizes the average number of local operations needed. We perform the local discrimination optimally in ascending order of $|<u_{i}|v_{i}>|$. Only if the former party obtains the error outcome, the successive party perform his local discrimination. It is explicit this protocol minimizes the average number of local operations.

In the paper \cite{Virmani}, optimal inconclusive discrimination of two non-orthogonal pure product states is also achieved by LOCC. This is also true in conclusive discrimination. These results may imply that two pure product multipartite states might have no non-local property although more than two can have \cite{Bennett}.

D. Yang thanks S. J. Gu and H. W. Wang for helpful discussion.
The work is supported by the NNSF of China (Grant No.19875041), the Special 
NSF of Zhejiang Province (Grant No.RC98022) and Guang-Biao Cao Foundation inZhejiang University.


\end{document}